\documentclass{aastex63}

\usepackage{booktabs}
\usepackage[utf8]{inputenc}
\usepackage{lineno} 
\usepackage{xcolor}
\usepackage{graphicx}

\received{month day, year}
\revised{month day, year}
\accepted{month day, year}

\shorttitle{Cooling process in G24 following burst}
\shortauthors{Xu et al.}

\begin{document}
%\linenumbers

\title{Cooling process of the high-mass young stellar object G24.33+0.14 following an accretion burst}

\correspondingauthor{Xi Chen}
\email{chenxi@gzhu.edu.cn}

\author[0009-0001-3487-1870]{Xiaoyun Xu}
\affiliation{Center for Astrophysics, Guangzhou University, Guangzhou 510006, People's Republic of China}

\author[0000-0002-5435-925X]{Xi Chen}
\affiliation{Center for Astrophysics, Guangzhou University, Guangzhou 510006, People's Republic of China}

\author[0000-0002-2895-7707]{Yang Yang}
\affiliation{College of Elementary Education, Changsha Normal University, Changsha 410100, People's Republic of China}



\begin{abstract}

The HMYSO G24.33+0.14 (G24), has recently been observed to undergo an accretion burst since September 2019, lasting approximately two years. By utilizing 1.3 mm observational data from the NOrthern Extended Millimeter Array (NOEMA) in March 2020 and the Atacama Large Millimeter/submillimeter Array (ALMA) in September 2019, we have examined the physical environment changes in gas and dust within G24 region during the decay phase of the accretion burst. Following the burst, the continuum emission in the inner core region of G24 diminished by approximately 20\%, while the emission in the outer region exhibited an increase by a factor of $\sim30\%$. This pattern indicates that the heat wave, triggered by the accretion burst, radiated outward from the core's interior to its periphery over the half-year period, with a calculated propagation speed of 0.08-0.38 times the speed of light. Moreover, the methanol emission intensity in this area has experienced a notable decline, with the rate of flux reduction correlating positively with the energy of the upper energy states. This, in conjunction with the analysis of methanol molecular line rotation temperature diagrams for different emitting regions, further substantiates that the core region of G24 cooled down, contrasted with the persistent heating in the outer region following the burst.

\end{abstract}


\section{Introduction}
\label{sec:intro}

The mass accretion rate of a star is not a constant throughout its formation process; instead, it predominantly exists in a state of low accretion rate, punctuated by brief episodes of accretion bursts characterized by a sudden spike in the rate of mass accretion. This intermittent pattern of mass accumulation is known as episodic accretion, as referenced in various studies \citep{2009ApJ...692..973E,2009ApJS..181..321E,2006ApJ...650..956V,2009ApJ...704..715V,2015ApJ...805..115V}.
Luminosity bursts detected in low-mass young stellar objects classified as FU-Orionis \citep{1966VA......8..109H} and EX-Orionis \citep{1989ESOC...33..233H} are believed to be indicative of these accretion bursts \citep{2014prpl.conf..387A,2015ApJ...805..115V}. These accretion bursts play a critical role in the mass accumulation phase of star formation, with the mass accreted during these bursts being responsible for at least 25\% of the star's total mass \citep{2019ApJ...872..183F}.

Recent observations have indeed confirmed that high-mass star formation experiences accretion bursts, similar to those observed in low-mass star formation. This phenomenon has been observed in high-mass young stellar objects (HMYSOs) such as S255IR NIRS3 \citep{2017NatPh..13..276C, 2018ApJ...863L..12L}, NGC 6334I-MM1 \citep{2017ApJ...837L..29H,2018ApJ...854..170H,2021ApJ...912L..17H}, and G358.93-0.03-MM1 \citep{2019ApJ...881L..39B,2020NatAs...4..506B}. Numerical simulations suggest that these accretion bursts are initiated by gravitational instabilities within the protostellar disk, leading to disk fragmentation and the formation of spiral arm structures. These spiral arms facilitate the rapid accretion of gas clumps onto the protostar, resulting in luminosity bursts \citep{2017MNRAS.464L..90M}. The presence of these spiral arm structures has been confirmed in actual observations, particularly in G358.93-0.03-MM1 \citep{2020NatAs...4.1170C,2023NatAs...7..557B}.
Furthermore, the  ``heat wave" propagation phenomenon, triggered by episodic bursts, has been observed in G358.93-0.03-MM1 \citep{2020NatAs...4..506B}. This heat wave which propagates outward from the inner to the outer regions of the molecular cores has also been deemed as a significant indicator of accretion burst event in the star formation process.
Studying these accretion burst events in HMYSOs provides critical insights into the formation processes of high-mass stars, highlighting the importance of episodic accretion in their development. 

To date, observations have captured accretion burst events in only a handful of HMYSOs. This limited number is largely attributed to the vast distances to regions where high-mass stars are forming, as well as the dense gas envelopes that enshroud HMYSOs, which make their detection a challenging task. However, the 6.7 GHz class II methanol maser has proven to be an invaluable tool for detecting such accretion bursts in HMYSOs. These masers are energized by intense infrared radiation fields. When an accretion burst occurs, the accretion luminosity increases, heating the surrounding dust and thus amplifying the infrared radiation. This enhanced radiation then excites the class II methanol masers, leading to a maser flare. It is worth noting that all five of the currently known HMYSOs with documented accretion burst events: S255IR-NIRS3, NGC6334I-MM1, G358-0.03-MM1, G323.46-0.08, and G24.33+0.14 were discovered through the detection of flares in their 6.7 GHz masers \citep{2017A&A...600L...8M,2018ApJ...854..170H,2019ATel12446....1S,2023A&A...671A.135K,2019MNRAS.487.2407P}.

\begin{figure}[!t]    
    \centering
    \includegraphics [width=0.7\textwidth]{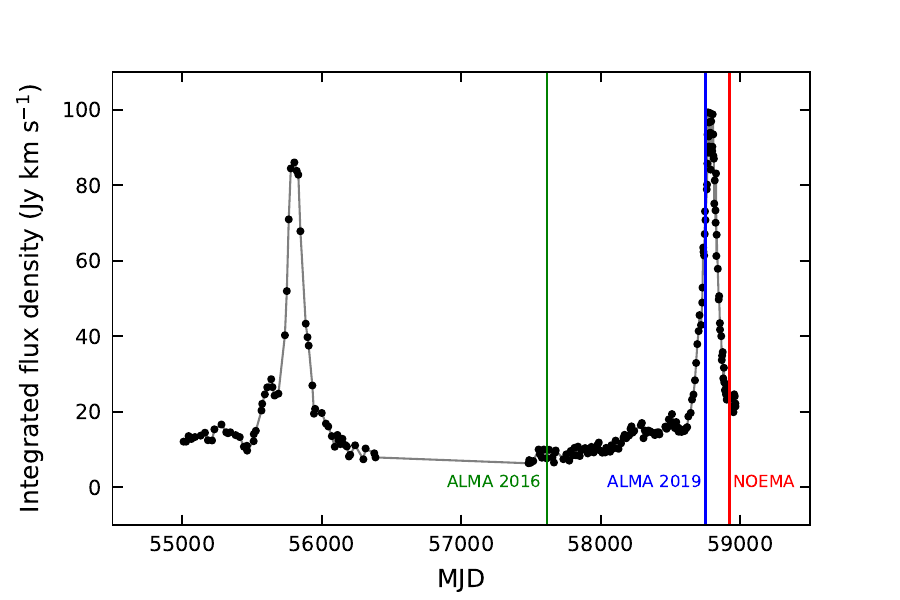}
    \caption{The 6.7 GHz methanol maser light curve for G24 \citep{2022PASJ...74.1234H}. The green, blue, and red vertical lines mark the ALMA observations in 2016 and 2019, and the NOEMA observation in 2020 at 1.3 mm, respectively.}
    \label{fig:1}
\end{figure}

Among the known maser burst sources, G24.33+0.14 (the target of this study, hereafter referred to as G24) is a HMYSO located at a distance of 7.20$\pm$0.76 kpc \citep{2022PASJ...74.1234H}, as determined using a Bayesian distance estimator  based on the Galactic rotation model of \cite{2019ApJ...885..131R}. 
Long-term monitoring of the 6.7 GHz methanol maser in G24 has revealed that, following a maser flare observed between November 2010 and January 2013 \citep{2018IAUS..336..319W}, another similar flare occurred in September 2019 \citep{2019ATel13080....1W}, as shown in Figure \ref{fig:1}. Combined with a similar luminosity burst behavior from the infrared observations, these findings suggest that G24 is a recurrent accretion burst source \citep{2022PASJ...74.1234H,2023ApJ...951L..24L}, exhibiting a cycle of 8.5 years and a burst duration of about 2 years. In addition to changes in infrared luminosity and methanol maser activity, significantly increased emission of both continuum and molecular spectral lines have also been detected in the millimeter band comparing the Atacama Large Millimeter/submillimeter Array (ALMA) data toward G24 between the quiescent in August 2016 and the burst stages in September 2019 \citep{2022PASJ...74.1234H}. However, the environmental changes surrounding G24 during the decay phase of the burst have not been previously examined. In this study, we have combined the 1.3 mm band observational data from NOrthern Extended Millimeter Array (NOEMA) in March 2020, which corresponds to the post-burst phase (Figure \ref{fig:1}), with ALMA data collected during the burst phase in September 2019. This comprehensive dataset allows us to investigate the changes in the continuum and molecular spectral lines of G24 throughout the post-burst stage, providing a more detailed understanding of the evolution of this HMYSO following its accretion burst.
We further enhance our analysis by incorporating data from ALMA observations of G24 during its quiescent phase in August 2016 (pre-burst period). This enables a more comprehensive understanding of gas temperature and density  variations in the G24 region both before and after the burst.

\section{Observation and data reduction} \label{sec:obs_data}

\subsection{Observation}
Interferometric observations of G24 during its post-burst phase were performed using the NOEMA in the 1.3 mm band, as part of Project ID: W19AP001 (P.I. Chen Xi). The key observational parameters are outlined in Table \ref{tab:obs parameter}. The target, G24, was centered at RA= $18^{\rm h}35^{\rm m}08\fs140$, Dec= $-07^{\circ}35'04''.15$ (J2000). Observations in the C array configuration were conducted with 10 antennas on the dates March 15, 25, 27, 28, and April 7, 2020, accumulating a total on-source integration time of 9.1 hours. The projected baselines ranged from 15.0 m (corresponding to 10.9 k$\lambda$) to 368 m (corresponding to 285.7 k$\lambda$). The lower sideband (LSB) and upper sideband (USB) covered frequencies from 213.3 GHz to 221.4 GHz and from 228.8 GHz to 236.9 GHz, respectively, with a channel width of 2 MHz, which corresponds to a velocity resolution of approximately 2.6 ${\rm km\, s^{-1}}$. 

The data were calibrated using the NOEMA pipeline within the the Continuum and Line Interferometer Calibration ($\mathtt{CLIC}$) software, which is a part of the Grenoble Image and Line Data Analysis Software (GILDAS)\footnote{\url{https://www.iram.fr/IRAMFR/GILDAS}}.  
Phase and amplitude calibration were performed using J1745-0753 and 1829-106. The flux calibrator was MWC349, and 3C345 was added as an additional flux calibrator on March 15, 2020. The absolute flux error is estimated to be less than 10\%. 
Bandpass calibration was carried out using the following sources: 3C279 (March 15, 2020), 1749+096 (March 25 and March 27, 2020), 3C84 (March 28, 2020), and 3C454.3 (April 7, 2020). 
The primary beam of the NOEMA antennas is $21.7''$ at 232.5 GHz.
Following calibration, the calibrated data (stored as uv table, where u and v represent the projected baseline lengths in units of observing wavelength) was generated and further processed using GILDAS/$\mathtt{MAPPING}$. Imaging of the uv tables was performed with natural weighting, and then deconvolved carried out the Högbom CLEAN algorithm \citep{1974A&AS...15..417H}. 
The beam size was $1.94'' \times 0.57''$ with a position angle (PA) of $10.88^{\circ}$, and $2.12'' \times 0.60''$ with a PA of $9.71^{\circ}$ at the USB and LSB, respectively.

\subsection{Data reduction for comparing with the ALMA data}

The ALMA 2016 data for the pre-burst stage of G24 were retrieved from the ALMA Cycle 3 archival data (Project code: 2015.1.01571.S), which was captured on August 15, 19 and 22, 2016.
The ALMA 2019 data for the burst stage of G24 were retrieved from the ALMA Cycle 6 archival data (Project code: 2018.A.00068.T), which was captured on September 25 and 26, 2019. A detailed description of the ALMA observations can be found in \cite{2022PASJ...74.1234H}. 
The ALMA 2016 observations encompassed three spectral windows (SPWs): SPW 0 spanned 212.9 – 213.8 GHz, SPW 1 from 216.5 – 217.5 GHz, and SPW 2 covered 229.5 – 230.0 GHz, with each channel having a width of 976.562 kHz.
The ALMA 2019 observations encompassed four SPWs: SPW 0 spanned 213.2 – 213.7 GHz, SPW 1 from 216.9 – 217.3 GHz, SPW 2 covered 229.5 – 230.0 GHz, and SPW 3 ranged from 230.1 – 230.6 GHz, with each channel having a width of 244.141 kHz.
For further specifics, refer to Table \ref{tab:obs parameter}. 

To assess the flux variations in the 1.3 mm band both during and following the burst, NOEMA data were extracted to correspond with the ALMA 2019 spectral windows using the $\mathtt{uv\_ extract}$ task within the GILDAS/$\mathtt{MAPPING}$ package, with the exception of SPW 0. Given that the NOEMA observations did not fully encompass ALMA 2019's SPW 0, the scope of the ALMA 2019 SPW 0 in this study was confined to the frequencies 213.3 – 213.7 GHz. Moreover, the rest frequencies of the NOEMA spectra were fine-tuned to coincide with those of the ALMA 2019 spectral windows. 
Subsequent processing of both NOEMA and ALMA data was carried out using the CASA (Common Astronomy Software Applications) software suite \citep{2007ASPC..376..127M}, version 6.5.6. The data were standardized to a consistent beam size of $2.12'' \times 0.61''$ at a PA of $10^{\circ}$, and a spectral resolution of 2 MHz was applied. The typical noise levels for NOEMA, ALMA 2019 and ALMA 2016 are 2.3 ${\rm mJy\, beam^{-1}}$, 3.0 ${\rm mJy\, beam^{-1}}$ and 5.0 ${\rm mJy\, beam^{-1}}$ for each channel, respectively.

\begin{table}[t!]
\centering
\caption{Observation Parameters of ALMA and NOEMA}
\label{tab:obs parameter}
\resizebox{\textwidth}{!}{
\begin{tabular}{lccc}
\toprule
Array & ALMA 2016 & ALMA 2019& NOEMA \\
\midrule
Project code & 2015.1.01571.S & 2018.A.00068.T & W19AP001 \\
Observation date & 2016 Aug 15, 19, and 22 & 2019 Sep 25 and 26 & 2020 Mar 15, 25, 27, 28, Apr 7 \\
On source time & 0.3 h & 1.6 h & 9.1 h \\
Baseline length & 15.1 m -- 1.5 km & 15.1 m -- 2.5 km & 15.0 m -- 368 m \\
Frequency cover & 
\begin{tabular}[c]{@{}c@{}} 
SPW 0: 212.9 -- 213.8 GHz \\ 
SPW 1: 216.5 -- 217.5 GHz \\ 
SPW 2: 229.4 -- 229.9 GHz  
\end{tabular}& 
\begin{tabular}[c]{@{}c@{}} 
SPW 0: 213.2 -- 213.7 GHz \\ 
SPW 1: 216.9 -- 217.3 GHz \\ 
SPW 2: 229.5 -- 230.0 GHz \\ 
SPW 3: 230.1 -- 230.6 GHz
\end{tabular} &
\begin{tabular}[c]{@{}c@{}} 
LSB: 213.3 -- 221.4 GHz  \\ 
USB: 228.8 -- 236.9 GHz 
\end{tabular} \\
Channel width & 976.562 kHz & 244.141 kHz & 2 MHz \\
Resolution & 
\begin{tabular}[c]{@{}c@{}} 
SPW 0: 0.29$''\times$0.25$''$, PA= $-80^\circ$ \\ 
SPW 1: 0.29$''\times$0.25$''$, PA= $-87^\circ$ \\ 
SPW 2: 0.27$''\times$0.24$''$, PA= $-83^\circ$ 
\end{tabular} & 
\begin{tabular}[c]{@{}c@{}} 
SPW 0: 0.24$''\times$0.14$''$, PA= $-74^\circ$ \\ 
SPW 1: 0.21$''\times$0.14$''$, PA= $-81^\circ$ \\ 
SPW 2: 0.20$''\times$0.13$''$, PA= $-81^\circ$ \\ 
SPW 3: 0.20$''\times$0.13$''$, PA= $-81^\circ$
\end{tabular} & 
\begin{tabular}[c]{@{}c@{}} 
LSB: 2.12$''\times$0.60$''$, PA= $10^\circ$ \\ 
USB: 1.94$''\times$0.57$''$, PA= $11^\circ$
\end{tabular} \\
\bottomrule
\end{tabular}
}
\end{table}

\section{Results} \label{sec:results}
\subsection{Continuum emission}
\label{subsec:contiuum}

\begin{figure}[!t]
    \centering
    \includegraphics [width=1.0\textwidth]{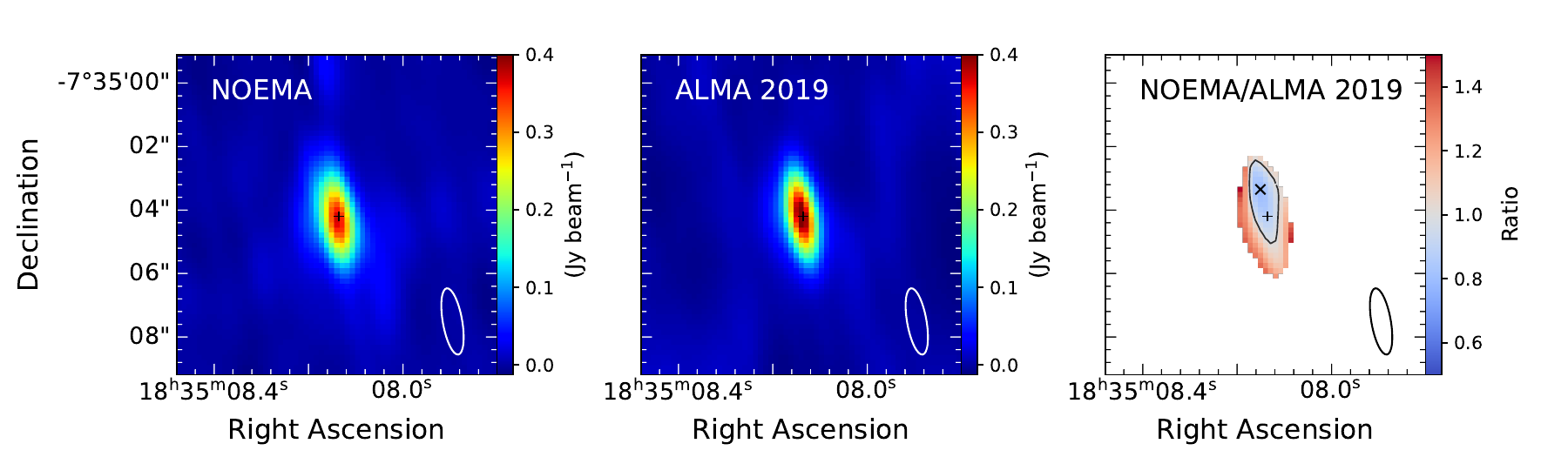}
    \caption{The averaged continuum images of NOEMA (left panel) and ALMA 2019 (middle panel) at the same frequency range, along with their flux ratio image (right panel).  The flux ratio was determined using signals exceeding 7 times the root mean square (rms) noise level in both NOEMA and ALMA 2019 images, with rms values of 5.4 $\rm{mJy\, beam^{-1}}$ and 7.9 $\rm{mJy\, beam^{-1}}$, respectively. The white or black ellipse in the lower right corner of each panel represents the beam size. In the right panel, the black cross denotes the center, RA= $18^{\rm h}35^{\rm m}08\fs136$, Dec= $-07^{\circ}35'04''.20$ (J2000), derived from 2D Gaussian fitting, while the $\times$ symbol marks the location of the minimum flux ratio. The black contour delineates a flux ratio of 1.}
    \label{fig:2}
\end{figure}

Figure \ref{fig:2} displays the averaged continuum images across the four SPWs, derived from both ALMA 2019 data collected in September 2019 and NOEMA data obtained in March 2020, presented at a uniform beam size, along with their corresponding flux ratio images. Given that the beam size used in this study is $2.12'' \times 0.61''$, which exceeds the $0.3''$ beam size utilized in \cite{2022PASJ...74.1234H}, the continuum images from this study reveal only a single continuum peak, as opposed to the three peaks observed by \cite{2022PASJ...74.1234H}. 

To investigate the variations in flux density of the continuum, we performed 2D Gaussian fitting on the continuum images of four different SPWs and averaged SPWs. The fitting results are presented in Table \ref{tab:continuum}. 
Within each SPW, the peak flux observed by NOEMA is consistently lower than that recorded by ALMA 2019. The mean ratio of the continuum peak flux, with NOEMA to ALMA 2019, is determined to be 0.87 $\pm$ 0.03, hinting at a decrease in the peak flux within the core region of G24 approximately 200 days after the maser flare. Despite being within the margin of error, the flux density ratio of NOEMA to ALMA 2019 exceeds 1 for SPW 1 and SPW 3, with an averaged continuum flux density ratio of 1.14$\pm$0.06, which is also larger 1. This suggests a marginal rise in the continuum flux density of G24 subsequent to the maser flare. However, this outcome seems contradictory to the observed decline in peak value, potentially attributable to enhanced emission from the core's periphery during the NOEMA observation period, as depicted in panel (c) of Figure \ref{fig:2}. We intend to delve deeper into this discrepancy in Section \ref{subsec:cont variability}.

\begin{table}
\caption{Properties of continuum emission from different spectral windows.}
\label{tab:continuum}
\resizebox{\textwidth}{!}{
\begin{tabular}{ccccccccc}
\toprule
Spectral window  & $F_{\rm NOEMA}$ & $F_{\rm 2019}$ &  $F_{\rm NOEMA}/F_{\rm 2019}$ & $I_{\rm NOEMA}$ & $I_{\rm 2019}$ & $I_{\rm NOEMA}/I_{\rm 2019}$ & Deconvolved size (NOEMA) & Deconvolved size (2019) \\
   & mJy & mJy  & ratio  & ${\rm mJy\, beam^{-1}}$ & ${\rm mJy\, beam^{-1}}$ & ratio & $'' \times \, ''$ & $'' \times \, ''$ \\
(1) & (2) & (3) & (4) & (5) & (6) & (7) & (8) & (9) \\
\midrule     
SPW 0 & 566(14) & 555(18) & 1.02(0.06) & 298(4) & 358(7) & 0.83(0.03) & 1.34(0.11)$\times$0.89(0.02) & 1.09(0.16)$\times$0.68(0.02) \\
SPW 1 & 627(15) & 573(20) & 1.09(0.06) & 332(4) & 368(8) & 0.91(0.03) & 1.34(0.11)$\times$0.89(0.03) & 1.02(0.18)$\times$0.70(0.03)\\
SPW 2 & 792(18) & 771(26) & 1.03(0.06) & 406(6) & 511(11) & 0.74(0.03) & 1.45(0.10)$\times$0.89(0.03)& 0.93(0.19)$\times$0.66(0.03)\\
SPW 3 & 833(19) & 742(26) & 1.12(0.07) & 429(6) & 493(11) & 0.87(0.02) & 1.45(0.10)$\times$0.88(0.03)& 0.93(0.19)$\times$0.66(0.03)\\
All SPWs & 709(16) & 621(19) & 1.14(0.06) & 369(5) & 426(8) & 0.87(0.03) & 1.40(0.10)$\times$0.89(0.03) & 0.86(0.17)$\times$0.62(0.02)\\

\bottomrule
\end{tabular}
}
\tablecomments{Column (1) lists the spectral windows. Columns (2) and (3) show the flux densities measured by NOEMA and ALMA 2019 from the 2D Gaussian fitting, respectively. Column (4) presents the ratio of the flux density of NOEMA to ALMA 2019. Columns (5) and (6) present the flux peaks of NOEMA and ALMA 2019 images, respectively. Column (7) displays the ratio of the flux peaks of NOEMA to ALMA 2019. Columns (8) and (9) list the deconvolved size of the 2D Gaussian fitting for NOEMA and ALMA 2019, respectively. Values in parentheses indicate the associated errors.}
\end{table}

\subsection{Methanol lines}
\label{subsec:methanol}
As the simplest complex organic molecule (COM) in interstellar space, methanol is widely distributed and serves as an excellent tracer for studying star formation. Beyond the prominent methanol masers, methanol thermal lines are also abundant and exhibit strong emission compared to other molecules. 

\begin{figure}[!t]    
    \centering
    \includegraphics [width=1.0\textwidth]{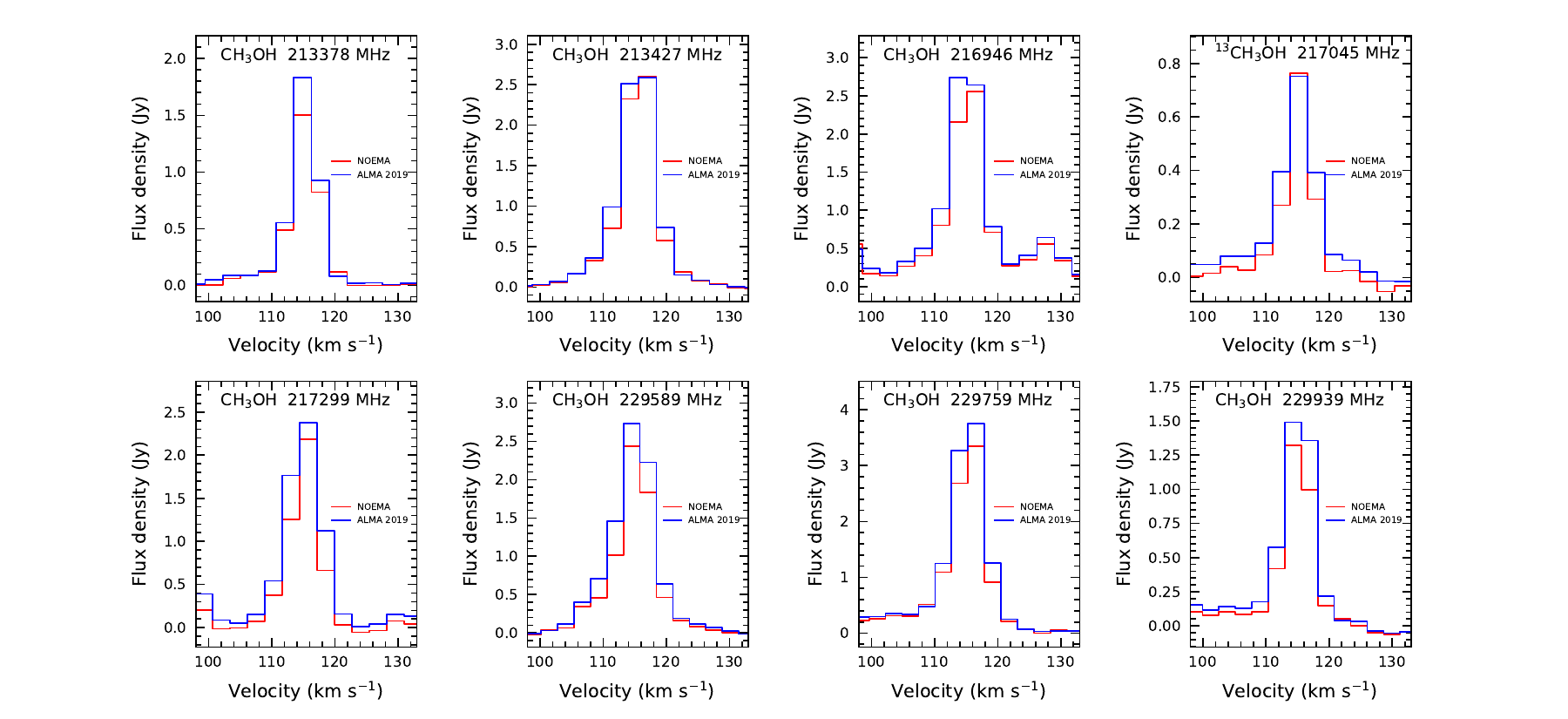}
    \caption{Methanol line spectra detected from the NOEMA and ALMA 2019 observations. 
    The red line represents NOEMA data, while the blue line corresponds to ALMA 2019 data. These spectral lines were extracted from an elliptical region, centered at the continuum peak position (coordinates RA = $18^{\rm h}35^{\rm m}08\fs136$ and Dec = $-07^{\circ}35'04''.20$; J2000), with a semi-axis 2.6$''\times1.3''$ at a PA of $10^{\circ}$.}
    \label{fig:3}
\end{figure}

This study focuses on the flux variations subsequent to the burst for eight methanol lines, including one $^{13}\rm{C}$ isotopologue and one torsionally excited line ($v_t=1$), as detailed in Table \ref{tab:methanol}. 
Figure \ref{fig:3} illustrates these methanol spectra, which were extracted from an elliptical region centered on the peak position of the continuum, with a semi-axis of 2.6$''\times1.3''$ oriented at a PA of 10 $^{\circ}$, from the observations conducted by NOEMA and ALMA 2019.
A visual inspection to the methanol line spectra from the two observations shows a decrease in the integrated intensity for all eight methanol lines by March 2020. Upon comparing the spectral peak flux densities between the two observations, it is observed that six methanol lines exhibit lower peak values in the March 2020 NOEMA data compared to the September 2019 ALMA data. Nevertheless, the NOEMA observations show peaks at 213.427 GHz and 217.045 GHz that are either similar to or comparable to those observed by ALMA 2019. For the line at 216.946 GHz, there is a one-channel discrepancy between the peak positions of NOEMA and ALMA 2019, potentially attributable to the spectral smoothing process.

\begin{figure}[!ht]    
    \begin{minipage}{0.95\linewidth}
        \centering
        \includegraphics[width=18cm]{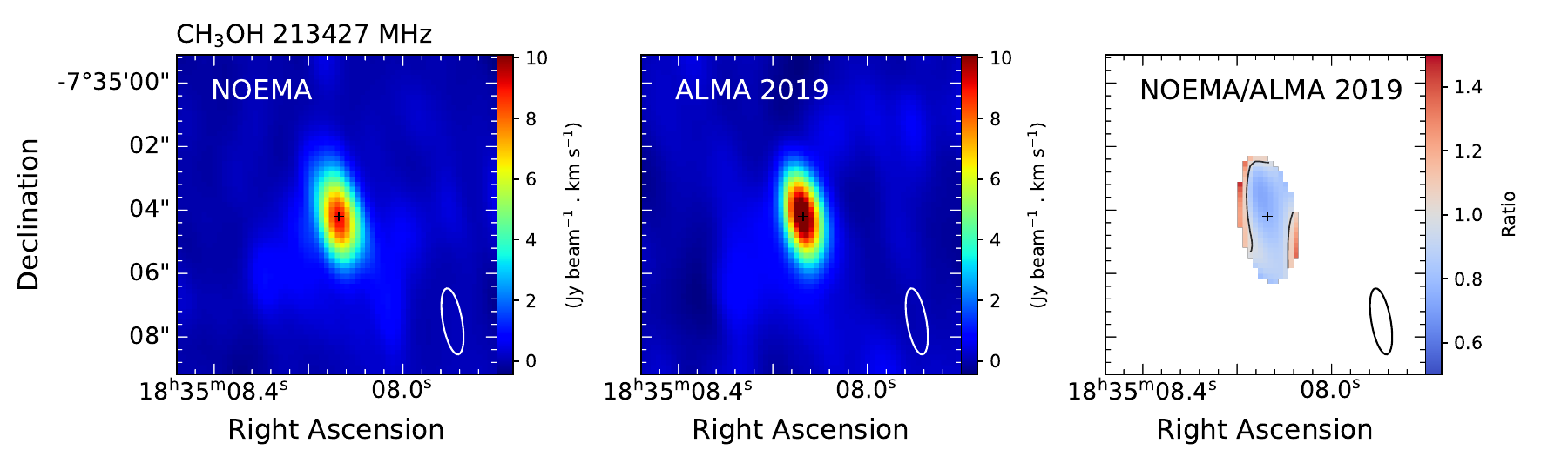}
    \end{minipage}
    \begin{minipage}{0.95\linewidth}
        \centering
        \includegraphics[width=18cm]{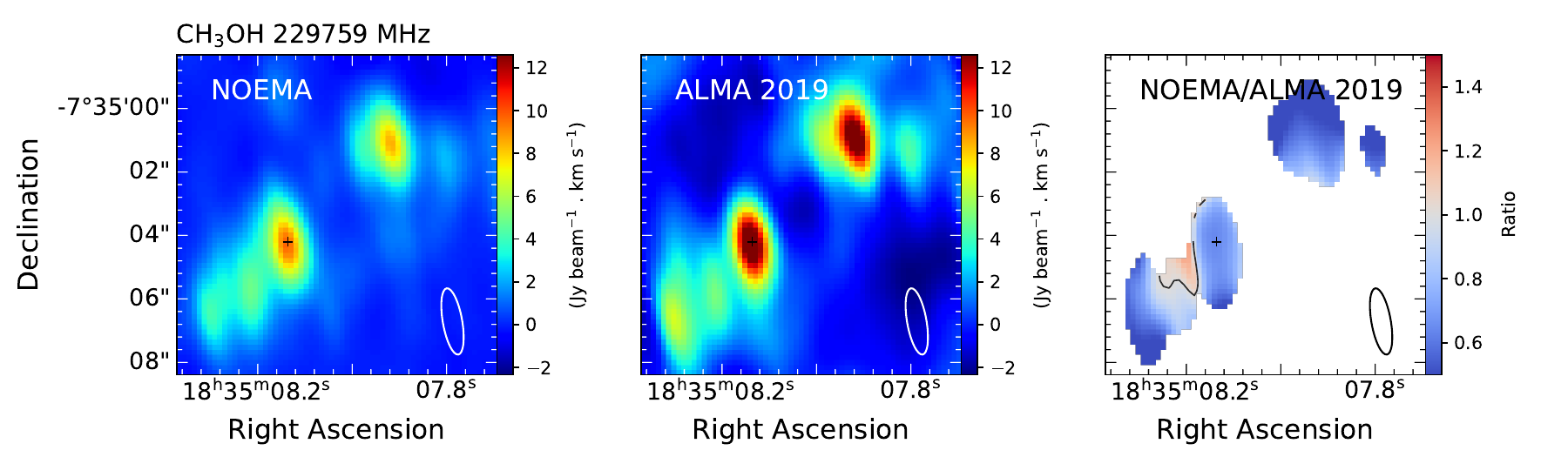}
    \end{minipage}
    \caption{Integrated intensity maps of methanol lines from NOEMA (left panel) and ALMA 2019 observations (middle panel), along with their flux ratio map (right panel). The 213.427 GHz and 229.759 GHz transitions are shown as examples. In addition to the 229.759 GHz transition, where the flux ratio is determined using signals exceeding 3 times the rms in both NOEMA and ALMA 2019 images, the flux ratios for all other transitions are calculated based on signals that are greater than 5 times the rms in the images. The white or black ellipse in the lower-right corner of each panel represents the beam size, and the black cross marks the central position obtained from the 2D Gaussian fitting. \\
    (The complete figure set (8 images) is available in the online journal.)}
    \label{fig:4}
\end{figure}

In order to gain a deeper understanding of methanol emission changes, the velocity range of [108,122] ${\rm km\, s^{-1}}$ was integrated to generate moment-0 maps for the methanol spectral lines. The NOEMA and ALMA 2019 integrated intensity maps, as well as their flux ratio maps, for the 213.427 GHz and 229.759 GHz lines are displayed as illustrative examples in Figure \ref{fig:4}. Additional maps for the other methanol lines are presented online. 
When comparing the ALMA 2019 data in September 2019 to the NOEMA data in March 2020, a substantial reduction in the integrated intensity of the 213.427 GHz line is observed. The 229.759 GHz line captures both the hot core region and the outflow, which is oriented in a northwest-southeast direction. The NOEMA integrated intensity map for the 229.759 GHz line indicates a significant drop in intensity within the hot core and outflow regions by March 2020. However, there is a minor increase in flux noted in the southeastern lobe adjacent to the hot core.

With the exception of the 229.759 GHz line, the integrated intensity maps for all other transitions reveal a single emission source. Given this work emphasis on flux variations within the core region, the integrated intensity of methanol emission within core region across the velocity range of [108, 122] $\rm{km\, s^{-1}}$ was determined by employing 2D Gaussian fitting. The results of the fitting are presented in Table \ref{tab:methanol}. Except for the 213.427 GHz line, the integrated intensity ratios of NOEMA to ALMA 2019 for all other methanol lines are below 1 within the margin of error. The weighted average flux density ratio is calculated to be 0.80$\pm$0.01.

\begin{table}
\caption{Parameters of the methanol transitions.}
\label{tab:methanol}
\resizebox{\textwidth}{!}{
\begin{tabular}{ccccccccccc}
\toprule
Molecule & Transition & Frequency & $E_{\rm u}/\kappa$ & $S\mu ^2$ & $F_{\rm NOEMA}$ & $F_{\rm 2019}$ & $F_{\rm NOEMA}/F_{\rm 2019}$ & Deconvolved size& rms (NOEMA) & rms (ALMA 2019) \\
      &       & MHz & K & D$^2$ &  ${\rm Jy. km\, s^{-1}}$ & ${\rm Jy. km\, s^{-1}}$ & ratio & $'' \times \, ''$ & ${\rm Jy\, beam^{-1}\cdot km\, s^{-1}}$ & ${\rm Jy\, beam^{-1}\cdot km\, s^{-1}}$ \\
(1) & (2) & (3) & (4) & (5) & (6) & (7) & (8) & (9) & (10) & (11)\\
\midrule
CH$_3$OH & $13_6-14_5$ E $v_t=0$ & 213377.528 & 389.92 & 10.18 & 8.61(0.17) & 10.23(0.17) & 0.84(0.03) & 0.98(0.11)$\times$0.71(0.02)& 0.065 & 0.082  \\
         & $1_1-0_0$ E $v_t=0$ & 213427.061 & 23.37 & 3.57 & 19.92(0.58) & 20.68(0.73) & 0.96(0.06) & 1.44(0.13)$\times$1.00(0.04) & 0.153 & 0.272\\
         & $5_4-4_2$ E $v_t=0$ & 216945.521 & 55.87 & 4.49 & 19.46(0.44) & 22.07(0.52) & 0.88(0.04) & 1.18(0.11)$\times$0.88(0.03) & 0.135 & 0.224\\
         & $6_1-7_2$ A$^-$ $v_t=1$ & 217299.205 & 373.92 & 18.56 & 12.45(0.25) & 16.79(0.26) & 0.74(0.03) & 1.02(0.11)$\times$0.75(0.03) & 0.092 & 0.121 \\
         & $15_4-16_3$ E $v_t=0$ & 229589.056 & 374.44 & 18.34 &16.83(0.33) & 22.58(0.94) & 0.75(0.05) & 1.08(0.10)$\times$0.73(0.02) & 0.124 & 0.460 \\
         & $8_{-1}-7_0$ E $v_t=0$ & 229758.756 & 89.10 & 20.19 & 18.78(0.75) & 21.81(2.43) & 0.86(0.13) & 1.20(0.11)$\times$0.87(0.03) & 0.204 & 1.04\\
         & $19_5-20_4$ A$^-$ $v_t=0$ & 229939.095 & 578.60 & 22.79 & 7.98(0.15) & 11.28(0.52) & 0.71(0.05) & 1.02(0.10)$\times$0.68(0.02) & 0.061 & 0.269 \\
$^{13}$CH$_3$OH & $14_1-13_2$ E $v_t=0$ & 217044.616 & 254.25 & 5.79 & 4.02(0.10) & 5.22(0.15) & 0.77(0.04) & 0.95(0.13)$\times$0.71(0.02) & 0.037 &0.075 \\
\bottomrule
\end{tabular}
}
\tablecomments{Columns (1) -- (5) are molecule name, transition, rest frequency and upper energy level, intrinsic line strength and magnetic dipole moment, respectively, taken from the Cologne Database for Molecular Spectroscopy (CDMS,  \citealt{2001A&A...370L..49M}). Columns (6) and (7) are the methanol integrated intensity obtained after 2D Gaussian fitting of NOEMA and ALMA 2019 integrated intensity maps, respectively. Column (8) is the ratio of the integrated intensity of NOEMA and ALMA 2019. Column (9) is deconvolved size obtained by fitting NOEMA data. Columns (10) and (11) are the rms noise level of integrated intensity maps of NOEMA and ALMA 2019, respectively.}
\end{table}

\section{Discussion} \label{sec:disc}
\subsection{Continuum variability}
\label{subsec:cont variability}

The flux ratio map of the NOEMA and ALMA 2019 continuum emission, as depicted in the right panel of Figure \ref{fig:2}, reveals that the continuum flux ratio is minimal in the vicinity of the central region and gradually increases with increasing distance from the core center. In the peripheral areas, the flux ratio even exceeds 1. This result suggests that at the heart of the core, the continuum emission detected by NOEMA in March 2020 is less intense compared to that captured by ALMA in September 2019. Conversely, in the outer sectors of the core, NOEMA's recorded emission surpasses ALMA 2019's.
The observed pattern suggests that the outer regions have undergone heating, while the inner regions have experienced cooling, likely due to the outward propagation of a heat wave triggered by accretion burst events, as elaborated in \cite{2020NatAs...4..506B}. This phenomenon is responsible for the observed decrease in NOEMA's continuum peak value and the increase in flux density, which are described in Section \ref{subsec:contiuum}.
We noticed that the minimum point of the continuum flux ratio is slightly offset from the fitted center. We believe this is due to the uneven distribution of dust and the different uv coverage of the interferometer arrays.

From September 25, 2019 to April 7, 2020, spanning 195 days, considering the distance range from the continuum fitting center to the location where the flux ratio equals 1 is between 0.36$^{\prime \prime}$ and 1.77$^{\prime \prime}$ (2600–13000 AU), we estimate the propagation speed of the heat wave in this region to be between 0.08 $c$ and 0.38 $c$.  
The variation in propagation speed may result from the inhomogeneous distribution of material in the region. 
In addition, we compared the deconvolved sizes obtained from fitting the average continuum images at the two observation epochs listed in Table \ref{tab:continuum}. The deconvolved size for ALMA 2019 is $0.86 '' \times 0.62''$, with an effective size of $0.73''$; for NOEMA, the deconvolved size is $1.40 '' \times 0.88''$, with an effective size of  $1.11''$. The effective size is calculated using the geometric mean  of the major and minor axes of the deconvolved size. The change in effective size is $0.38''$, corresponding to a heat wave propagation speed of 0.08c, which falls within the above recalculated range of heat wave propagation speeds.

It should be noticed that the heat wave propagation velocity estimated by \cite{2023A&A...671A.135K} based on the 6.7 GHz methanol maser (0.33 $c$) falls within the range of our estimated speeds.  \cite{2023A&A...671A.135K} have reported that methanol masers are predominantly located within the inner regions of the rotating disk. However, the majority of the 1.3 mm dust continuum emission is found to originate from more extended regions. Time-dependent radiative modeling of dust continuum emission during an MYSO accretion burst \citep{2024A&A...688A...8W} indicates that high optical depths significantly impede the transfer of energy. Given these findings, it is somewhat surprising that the propagation speeds of the heat wave in the regions of methanol masers and 1.3 mm continuum emission appear to be comparable. One plausible explanation might be that the actual distribution of dust is more complex than that depicted in the canonical disk/envelope model.

\subsection{Methanol variability}
\label{subsec: methanol variability}

\begin{figure}[!ht]    
    \centering
    \includegraphics [width=0.9\textwidth]{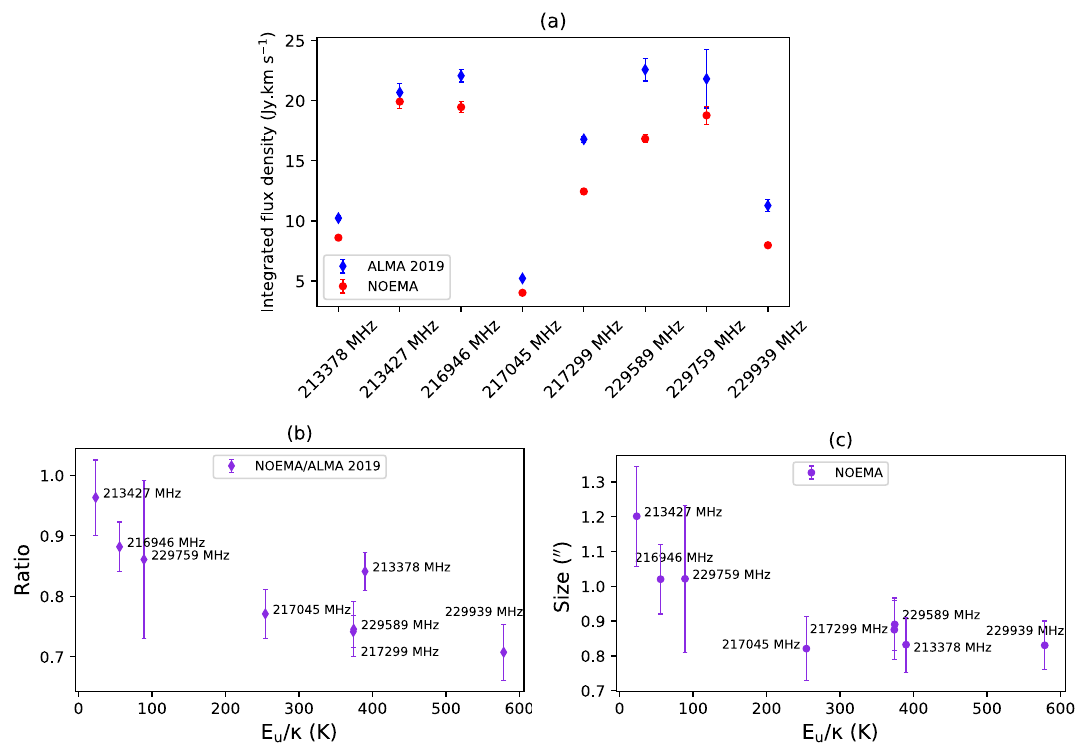}
    \caption{A comparative analysis of the integrated intensities of eight methanol lines, as detected by ALMA 2019 and NOEMA. Panel (a) displays ALMA 2019 data, symbolized by blue diamonds, in contrast to NOEMA data, indicated by red circles. Panel (b) depicts the ratio of integrated intensities for the methanol lines, with NOEMA values relative to those from ALMA 2019. Notably, for the 229.759 GHz line, which is characterized by an outflow structure, the comparison is restricted to the flux emanating from the central hot core region. Panel (c) shows the sizes of the emission regions for different methanol lines observed by NOEMA.}
    \label{fig:5}
\end{figure}

The panel (a) of Figure \ref{fig:5} presents a graphical depiction of the integrated intensity for eight methanol lines, observed with both NOEMA and ALMA 2019, as extracted from 2D Gaussian fits. As shown in the panel (b) of Figure \ref{fig:5}, the integrated intensity ratio of NOEMA to ALMA 2019 for these methanol lines decreases as the energy of the upper energy state increases. This suggests that lines with higher upper state energies undergo more significant flux attenuation. In contrast, during the burst phase, \cite{2022PASJ...74.1234H} noted an opposing trend, where not only methanol but also 16 lines from four other oxygen-bearing COMs exhibited an increasing trend in flux ratios with the increasing energy of the upper energy state.

\cite{2022ApJS..260...34Z} investigated the variability of W51-IRS2 under the assumption that varying ranges of ammonia's upper state energies are indicative of different core layers, with the higher energy states being associated with the more interior layers.
Building on this premise, our current study infers that the more significant reduction in flux density observed in methanol lines with higher upper energy states, as depicted in the panel (b) of Figure \ref{fig:5}, could suggest a temperature decrease or cooling process within the core's inner regions. To further probe this phenomenon, we examine the emitting region scales of various methanol lines.
The effective size of methanol emission is calculated using the the geometric mean of the major and minor axes of the deconvolved size of the methanol component listed in column (9) of Table \ref{tab:methanol}. As illustrated in panel (c) of Figure \ref{fig:5}, among the eight methanol lines, those with lower upper state energies (below 100 K) display larger emission regions, while those with higher upper state energies exhibit smaller regions. This observation lends partial support to the hypothesis that transitions with higher upper state energies are typically generated in the more central regions of the core. If this is indeed the case, the anti-correlation observed between the upper state energies and the ratio of integrated intensities suggests that the cooling process is likely a progressive one, starting from the core's innermost layers and moving outward. This pattern of cooling from the inside out mirrors the propagation of heat waves from the deepest regions to the outer layers, revealing the intricate dynamics of heat transfer within the core.

\subsection{Rotational temperature diagram analysis for methanol}
\label{subsec: rotational temperature diagram}

Given the observed phenomenon of central cooling and peripheral heating, presumably due to the heat wave propagation, as evidenced in variability of both continuum and methanol line emission discussed above, we have conducted a more in-depth analysis of the physical condition variations across various regions surrounding G24 at the ALMA 2016, ALMA 2019 and NOEMA observation epochs. Based on the assumption of local thermodynamic equilibrium (LTE), the rotational temperature diagram analysis can be used to estimate the rotational temperature and column density of the methanol molecule. The following formula is used for the analysis of the rotational temperature diagram \citep{2013ApJS..206...22C}:
\begin{equation}
\ln\left(\frac{3\kappa W}{8\pi^3\nu S\mu^2}\right)=\ln\frac{N(T_{\mathrm{rot}}-T_{\mathrm{bg}})}{QT_{\mathrm{rot}}}-\frac{E_u/\kappa}{T_{\mathrm{rot}}}\, ,
\label{eq:T_rot}
\end{equation}
where $\kappa$ is the Boltzmann constant, $W$ is the integrated intensity, $\nu$ is the spectral line frequency, $S$ is the intrinsic line strength, $\mu$ is the dipole moment, $N$ is the column density, $Q$ is the partition function, $T_{\rm rot}$ is the rotational temperature, $T_{\rm bg}$ is the microwave background radiation, $E_u/\kappa$ is the upper state energy level. 
The partition function for ${\rm CH_3OH}$ \citep{2013ApJS..206...22C} is 
\begin{equation}
    Q(T_{\rm rot})=1.2327 T^{1.5}_{\rm rot}\, .
\end{equation}

\begin{figure}[!t]
    \centering
    \includegraphics [width=1.0\textwidth]{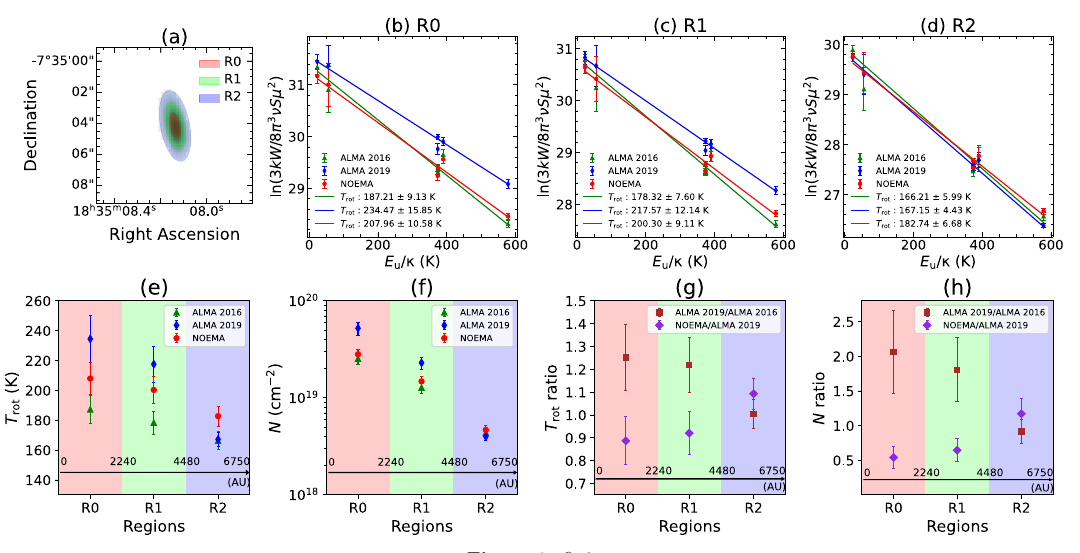}
    \caption{Panel (a): Schematic of the ring regions used for rotational temperature diagram analysis. Panels (b) -- (d) show the rotational temperature diagrams for R0, R1 and R2 regions, respectively. The solid lines in the plots represent the fitting lines (green for the ALMA 2016 data, blue for the ALMA 2019 data, and red for the NOEMA data), which were derived using only six methanol lines, excluding the 229.759 GHz line. Panels (e) through (h) display the rotational temperature, column density, and the ratios of rotational temperatures and methanol column densities between NOEMA and ALMA 2019 data, as well as between ALMA 2019 and ALMA 2016 for regions R0 to R2, respectively. The spatial scales, measured in AU, are based on the semi-minor axis of the elliptical rings depicted in panel (a). }
    \label{fig:6}
\end{figure}

We have delineated three specific regions -- R0, R1, and R2 --  ranging from the inner to the outer zones of the methanol emission area for an in-depth rotational temperature diagram analysis, as illustrated in panel (a) of Figure \ref{fig:6}. The rotational temperature diagrams with the derived rotational temperatures and column densities for these regions from NOEMA and ALMA data are shown in panels (b) -- (d) of Figure \ref{fig:6}. These measurements provide valuable insights into the thermal and compositional structures of the methanol molecule across different regions surrounding G24, following the burst event. The derived parameters exhibit consistent decrease trends with the increasing distance from the core center at three observation epochs, indicating rotational temperature diagram analysis yields reliable outcomes.

By analyzing the parameters collected over three different epochs, it is evident that the gas temperature and methanol column density in the inner layers (R0 and R1) observed by NOEMA and ALMA 2016 are generally lower than those observed by ALMA 2019. Furthermore, the gas temperature and methanol column density observed by ALMA 2016 are lower than those observed by NOEMA. This trend contrasts sharply with the outer layer (R2). In the outer layer, the two parameters of ALMA 2016 data are basically consistent with ALMA 2019 data, while both parameters increased in the NOEMA data, as illustrated in panels (e) and (f) of Figure \ref{fig:6}. Panels (g) and (h) of Figure \ref{fig:6} further indicate that the ratio of gas temperature and column density between ALMA 2019 and ALMA 2016 exhibits a decreasing trend from the inner to the outer layers. In the inner layers (R0 and R1), the ratio is greater than 1, indicating higher values in ALMA 2019 data. Conversely, in the outer layer (R2), the ratio is approximately equal to 1, suggesting that the heating effect had not yet propagated to this region during the burst phase of G24. In contrast, the ratio of gas temperature and column density between NOEMA and ALMA 2019 displays an increasing trend across the three layers, from the innermost (R0) to the outermost (R2).
In the inner layers R0 and R1, the ratios are below 1, whereas in the outer layer R2, they surpass 1. This outcome is consistent with the analysis of the continuum flux ratio maps detailed in Section \ref{subsec:cont variability}, which suggests that the inner regions of G24 are experiencing a cooling effect, whereas the outer regions are heated over the half-year span separating the NOEMA and ALMA 2019 observations. 
Given that the ALMA 2016 observation period represents the initial state of G24, we can analyze the timescale for the temperature and column density in the R0 region to return to their initial states following the 2019 burst. Assuming a uniform decrease in column density from the 2019 burst to March 2020 at a rate of $r_{N} = \Delta N / \Delta t$, where $\Delta N$ is the change in column density and $\Delta t$ is the time separation, we can estimate the timescale for the methanol in the R0 region to cool down to its initial state.
Using the ALMA 2019 and NOEMA data, it is derived that this timescale is approximately $218 \pm 109$ days. Considering the minimum propagation speed of the heat wave in this region (0.08 $c$), the heat wave would take approximately 500 days to leave the hot core region after the burst. Consequently, the hot core could entirely cool down to its initial state about 700 days after the burst.
Based on these observations, we can infer a sequence of events following the burst: an initial phase characterized by a gradual increase in temperature, followed by a cooling phase that propagates from the central regions outward into the more distant areas as the system evolves post-burst at a typical timescale of $\sim700$ days. This sequence provides a comprehensive view of the dynamic thermal changes within G24, highlighting the complex interplay between heating and cooling in the aftermath of such energetic events.

The reason for the changes in methanol column density following the burst is unclear. It may be because the temperature rises, causing the methanol ice on dust grains to sublimate and increasing the amount of methanol in the gas phase. As shown in panel (f) of Figure \ref{fig:6}, in March 2020 after the burst, the column density in the inner region decreased significantly, nearing the level before the burst in August 2016. Based on the estimated cooling timescale, the entire hot core is expected to return to the 2016 state approximately 700 days after the burst. According to the simulation results by \cite{2024A&A...684A..51G} on the variation of methanol abundance in high-mass stars during the pre-burst, burst, and post-burst phases, the timescale for methanol gas to refreeze after a burst is about 100 years. This implies that for sources like G24, which experience repeated bursts within short intervals, the abundance of gaseous methanol should continue to increase. However, this does not match our observational results.
We believe that the change in methanol column density is more likely caused by variations in local temperature in G24, which alter the proportion of excited methanol gas. The propagation of the heat wave raises the temperature in regions that were previously below the methanol excitation threshold, leading to stimulated methanol emission and increased flux, which appears as an increase in column density. Temperature variations significantly affect the behavior of molecules and the overall physical state of the protostellar cloud.

\section{Summary} \label{sec:sum}
By analyzing the 1.3 mm band continuum and methanol lines in the vicinity of the episodic burst HMYSO G24, as captured by ALMA in August 2016 during the quiescent stage, in September 2019 during the burst phase and by NOEMA in March 2020 during the post-burst phase, we have arrived at the following conclusions:

1. In comparison to the ALMA 2019 observations taken during the burst phase, the continuum images of G24 from the NOEMA data indicate a cooling within the inner regions of the continuum emission, while the outer regions remain heated following the burst. We deduce that the sustained heating in the outer regions is likely caused by a heat wave triggered by the episodic accretion event, with an estimated propagation speed of approximately $0.08\,c - 0.38\,c$. The value previously estimated based on the 6.7 GHz methanol maser observations also falls within this range.

2. All eight methanol lines studied in this work, with the exception of the 213.427 GHz line, have shown a marked decrease in methanol line integrated intensity from September 2019 to March 2020. Furthermore, there is a correlation between the magnitude of the integrated intensity reduction and the energy level of the upper state: as the energy of the upper state increases, the decrease in intensity becomes more pronounced.

3. The analysis of the methanol rotational temperature diagram shows that by March 2020, there was a modest decrease in both the gas temperature and column density of methanol in the innermost layer of the methanol emission region, as opposed to a rise in these measurements in the outer layer when compared with the figures from September 2019. This trend implies that both the temperature ratio and column density ratio between the two observation periods generally increased with increasing distance from the core, with the outermost layer even surpassing a ratio of 1. 

Items 2 and 3 provide additional support to the proposition outlined in item 1 concerning the propagation of a heat wave through the dust and molecular core following the burst. This process results in an initial phase of gradual heating, succeeded by a cooling effect, which transitions from the innermost regions to the outer areas in the aftermath of the burst. Collectively, these observational findings reinforce the hypothesis that the physical environment variations observed are a direct consequence of the bursts that occur during the high-mass star formation process.

\acknowledgments
The authors kindly thank the anonymous referee for the comments that helped improve the manuscript. This work is supported by the National Key R\&D Program of China (2022YFA1603102) and the National Natural Science Foundation of China (12473024). X.C. thanks Guangdong Province Universities and Colleges Pearl River Scholar Funded Scheme (2019). This work makes use of the following ALMA data: ADS/JAO.ALMA\#2015.1.01571.S, ADS/JAO.ALMA\#2018.A.00068.T. ALMA is a partnership of ESO (representing its member states), NSF (USA) and NINS (Japan), together with NRC (Canada), NSTC and ASIAA (Taiwan), and KASI (Republic of Korea), in cooperation with the Republic of Chile. The Joint ALMA Observatory is operated by ESO, AUI/NRAO and NAOJ.

\vspace{5mm}

\bibliography{G24}{}
\bibliographystyle{aasjournal}



\end{document}